\documentclass[aps,prd,showkeys,groupedaddress,longbibliography,nofootinbib,preprint]{revtex4-1}
\usepackage{mathrsfs,amsmath,amssymb,amsfonts}
\usepackage{graphicx}
\usepackage[colorlinks=true]{hyperref}
\usepackage{color}
\usepackage{xcolor}

\newcommand{\bw}{\begin{widetext}}
\newcommand{\ew}{\end{widetext}}
\newcommand{\be}{\begin{equation}}
\newcommand{\en}{\end{equation}}
\newcommand{\bea}{\begin{eqnarray}}
\newcommand{\ena}{\end{eqnarray}}

\begin{document}

\title{Influences of the coordinate dependent noncommutative space on charged and spin currents}

\author{Ya-Jie Ren}
\author{Kai Ma}
\email{makainca@yeah.net}

\affiliation{Department of Physics Science, Shaanxi University of Technology, Hanzhong, 723001, People's Republic of China}

\begin{abstract}
We study the charged and spin currents on a coordinate dependent noncommutative space. Starting from the noncommutative extended relativistic equation of motion, the non-relativistic approximation is obtained by using the Foldy-Wouthuysen transformation, and then the charged and spin currents are derived by using the extended Drude model. We find that the charged current is twisted by modifying the off-diagonal elements of the Hall conductivity, however, the spin current is not affected up to leading order of the noncommutative parameter. 
\end{abstract}



\date{\today}

\maketitle


\section{Introduction}\label{intro}
Nontrivial algebras of the coordinate operators in quantum mechanics had been considered to be important in describing the dynamics of physical systems at both low and high energy scale. It was used to describe the non-relativistic dynamics of a charged particle in a strong magnetic field under a perturbation of external electric field~\cite{Peierls1933}. A Lorentz invariant noncommutative space-time was also proposed in Ref.~\cite{Snyder:1946qz,Yang:1947ud}. Even though it was point out that extra dimensions are necessary to preserve the translation symmetries~\cite{Yang:1947ud}, theoretical studies became increasingly extensive. In Ref.~\cite{Snyder:1946qz}, the commutator between the coordinates is specified by the generators of Lorentz group, and hence its algebra is not closed by only the coordinate operators. Correspondingly, the product of the ordinary functions is twisted, and generally depends on their derivatives. Furthermore, by the uncertainty principle, the nonzero commutators between the coordinate operators implies that there is a minimal length, $\ell_{G}$, and hence noncommutative space-time was suggested to regularize the ultraviolet divergences in quantum field theory.

The most extensively studied noncommutative space-time is described by the algebra $[X_{\mu}, X_{\nu}] = i\theta_{\mu\nu}$, where $\theta_{\mu\nu}$ a totally anti-symmetric constant tensor~\cite{Chaichian:2000si, Zhang:2004yu,Ma:2011gc,Basu:2012td,Ma:2014tua,Wang:2015cua,Wang:2013kp,Deriglazov:2016mhk,Wang:2017azq,Wang:2017arq,Ma:2016rvw,Deriglazov:2015zta,Deriglazov:2015wde,Ma:2016rhk,Ma:2016vac,Das:2016hmc,Ma:2017fnt,Ma:2017rwg,Ma:2018sts}. 
The motivation for those works is that in the low energy approximation of string theory, the coordinate operators can be effectively described by the above algebra~\cite{Seiberg:1999vs}. Furthermore, it can also appear in the effective theory of quantum gravity~\cite{Freidel:2005me, Moffat:2000gr, Moffat:2000fv, Faizal:2013ioa}. However, due to that $\theta_{\mu\nu}$ is assumed to constant over the whole space-time, the rotational symmetry is explicitly broken in these effective theory~\cite{Douglas:2001ba, Szabo:2001kg}.

Recently, a general form of coordinate-dependent noncommutative algebra was proposed in Ref.~\cite{Kupriyanov:2013jka},
\begin{equation}\label{eq:nccAlgebra}
\big[ X_{\mu}, X_{\nu} \big] = i\theta \,\omega_{\mu\nu}(X) \,,
\end{equation}
where the noncommutative parameter $\theta=\ell_{G}^{2}$; the dimensionless operator $\omega_{\mu\nu}(X)$ is defined from physical considerations. Its possible expressions are further restricted by the Jacobi identity, $\omega^{\rho\nu}\partial_{\nu}\omega^{\sigma\lambda} + \omega^{\lambda\nu}\partial_{\nu}\omega^{\rho\sigma}+\omega^{\sigma\nu}\partial_{\nu}\omega^{\lambda\rho} = 0$, where $\omega^{\rho\nu}(x)$ is the corresponding dimensionless field of the operator $\omega_{\mu\nu} (X)$ with specified ordering.
Rotational properties and twisted interactions of this kind of noncommutative space-time have been studied in Ref.~\cite{Falomir:2015una} and and Ref.~\cite{Hassanabadi:2015tra}. A more general formulation of the noncommutative space-time has been studied in Ref.~\cite{Meljanac:2016jwk}. Non-zero components $\omega_{0i}$ can potentially induce non-unitarity problems for a fundamental theory. However, for phenomenological studies formal perturbative expansion is sufficient. For convergence and non-perturbative consequences of the underlying noncommutative space-time, we refer to the Refs.~\cite{Freidel:2005me,Steinacker:2011wb,Steinacker:2012ra,Vassilevich:2013ai,Rosa:2012pr,Galikova:2013jda}.

In this paper, we study the deformations on the charged and spin currents by the noncommutative algebra \eqref{eq:nccAlgebra}, which can be realized by using following $\star$-product (up to leading order of the noncommutative parameter $\theta$)~\cite{Kupriyanov:2013jka},
\begin{equation}
f(x) \star g(x) = f(x) g(x) + \frac{i}{2}\theta\omega^{\alpha\beta}(\partial_{\alpha} f)(\partial_{\beta}g)(x)\,.
\end{equation}
We will use this approach to study the noncommutative corrections. Furthermore, the coordinate operators $X$ are required to be self-adjoint with respect to the scalar inner product with multiplication rules defined by the $\star$-product~[30]. Similarly, the momentum operators $P$ are also required to be self-adjoint. Under these conditions, the momentum components are commutative with each others, {\it i.e.}, $[P^{\mu},\,P^{\nu}]=0$. However, the standard Heisenberg algebra is affected, and up to leading order of $\theta$ the deformed algebra reads $[X^{\beta},\,P^{\nu}] = i\eta^{\beta\nu} - \frac{i}{2}\theta \big( \partial^{\nu} \omega^{\beta\alpha}P_{\alpha}
+ \frac{i}{2} \partial_{\nu} \big[ \omega^{\beta\alpha} \partial_{\alpha}\ln\mu \big] \big)$~\cite{Kupriyanov:2013jka}. For a spinor particle with mass $m$ and charge $eQ$ in the external electromagnetic field defined by the vector potential $A_{\mu}(x)$, the deformed Dirac equation is given as~\cite{Kupriyanov:2013jka},
\begin{equation}
\gamma^{\mu} \big[ p_{\mu} - \partial_{\mu}\ln\mu(x) - e A_{\mu}(x) \big] \star \psi(x) - m \psi(x) =0\,,
\end{equation}
where the function $\mu(x)$ is defined as the measure for square-integrability and satisfies the equation $\partial_{\alpha}(\mu\omega^{\alpha\beta})=0$. 

The contents of this paper are organized as follows: in Sec.\ref{SOInter}, we will study the noncommutative corrections on the non-relativistic Hamiltonian for a spinor charged particle under a perturbation of an external electric feild. In Sec.\ref{NCSHalleff}, based on the extended Drude model, we will calculate the charged and spin dependent electric currents, as well as their corresponding conductivities. Our conclusions are given in
the final Sec.\ref{concl}.

\section{Spin-Orbital Interaction}\label{SOInter}
In this paper, we will study the noncommutative space-time with the operator $\omega^{\alpha\beta}$ is given as follows,
\begin{equation}\label{eq:ncOmega}
\omega^{\alpha\beta}(x) = \frac{1}{ \sqrt{\theta} } \epsilon^{\alpha\beta\mu} x_{\mu}\,,
\end{equation}
where $\epsilon^{\alpha\beta\mu}$ is the totally anti-symmetric tensor. In this case the function $\mu(x)=1$~\cite{Kupriyanov:2013jka}. As a result, the deformed Dirac equation reduce to
\begin{equation}\label{eq:nccDirac:reduce}
\gamma^{\mu} \big[ p_{\mu} - eA_{\mu}(x) \big] \star \psi(x) - m \psi(x) =0\,.
\end{equation}
Up to leading order of the noncommutative parameter, the deformed Dirac equation equation can be written as,
\begin{equation}\label{eq:nccDirac:star}
\gamma^{\mu} (p_{\mu} \psi - eA_{\mu}) \psi - \frac{\theta}{2}e \omega^{\alpha\beta}(\partial_{\alpha}A_{\mu}) \gamma^{\mu} (p_{\beta}\psi)  - m \psi =0\,.
\end{equation}
As expected the additional interaction depends on the momentum of the charged particle, and hence non-local. In next section, we will study the non commutative corrections on the spin-orbital interaction.

To study its non-relativistic charged and spin current, we take the non-relativistic approximation of the equation \eqref{eq:nccDirac:reduce} to obtain the corresponding minimal and spin-orbital interactions. It is well known that the approxiamtion can be addressed by using the Foldy-Wouthuysen unitary transformation~\cite{PhysRev.78.29}, which block diagonalizes the Dirac Hamiltonian by separating the positive and negative energy part of its spectrum.

We will study the charged and spin currents generated by external electric field via the minimal and spin-orbital interactions, respectively, hence the corresponding Dirac Hamiltonian of \eqref{eq:nccDirac:star} is,
\begin{equation}\label{HD}
H=c\,\vec{\alpha}\cdot \vec{p} + \beta mc^2 +
eV(\vec{x}) + eU(\vec{x}, \vec{p}),
\end{equation}
where $\beta=\gamma_0$ and $\alpha_i =\gamma_0\gamma_i$ are the Dirac matrices; $\vec p=-i\hbar\vec\bigtriangledown$ is the canonical momentum operators; $U(\vec{x}, \vec{p})$ is the noncommutative corrected scalar potential and given as follows,
\begin{equation} 
U(\vec{x}, \vec{p}) 
= 
- \frac{ 1 }{2} \sqrt{\theta} \, ( \vec{\nabla} V )\, \cdot \vec{L}\,,
\end{equation}
for the noncommutative algebra with $\omega_{\alpha\beta}$ described by \eqref{eq:ncOmega}, and $\vec{L}$ is the angular momentum operator. 

By the unitary Foldy-Wouthuysen transformation~\cite{PhysRev.78.29}, the nonrelativistic Hamiltonian is given as,
\begin{equation}\label{CSFWH}
    H_{FW}=\beta\bigg(mc^2+\frac{O^2}{2m}
    \bigg)+ e \epsilon-\frac{ e }{8m^2c^2}[O,[O,
    \epsilon]] \,,
\end{equation}
where,
\begin{eqnarray}\label{EEM}
    O &=& \vec{\alpha} \cdot \vec{p} \,,
    \\[3mm]
    \epsilon &=&  V(\vec{x}) - U(\vec{x}, \vec{p}) \,.
\end{eqnarray}
It is easy to get that $O^{2} =  \vec{p}^{2}$, and 
\begin{equation}\label{EEM}
    [O,[O,\epsilon]] 
    = \hbar^{2}\vec{\nabla}\cdot\vec{E}
    +2\hbar\vec{\Sigma}\cdot(\vec{E}\times\vec{p}) \,.
\end{equation}
Here we have neglected the term proportional to $\vec{\nabla}\times\vec{E}$. Furthermore, the terms proportional to $\vec{p}^{2}$ are also neglected, because their contribution is scaled by the factor $\sqrt{\theta} E/ m$ which is very small in practice (for example, it is of order 1 only when the external electric field is of order $10^{6}\rm{V}$ for free electron.). The term proportional to $\vec{\nabla}\cdot\vec{E}$ gives a constant shift of the energy in case of that the charged density is a constant (this is our case, see below), and hence is not physical. Under those approximation, the non-relativistic Hamiltonian is
\begin{equation}\label{eq:nc:NonRel}
    H=\frac{\vec{p}^2}{2m} -\frac{ e\hbar }{4m^2c^2} \vec{\sigma}\cdot(\vec{E}\times\vec{p}) + \sqrt{\theta}  H_{\theta}\,,
\end{equation}
with
\begin{equation}\label{eq:nc:NonRel}
    H_{\theta}=\frac{ 1 }{2} e \, \vec{E} \, \cdot \vec{L} \,.
\end{equation}
One can see that, the noncommutative algebra \eqref{eq:nccAlgebra} with $\omega_{\alpha\beta}$ described by \eqref{eq:ncOmega}, up to leading order of the noncommutative parameter, the Hamiltonian receives only spin-independent noncommutative correction. The additional term depends on the angular momentum of the matter particle. We will study the charged and spin currents, as well as their conductivities in next section.

\section{Spin Current and Spin-Hall Conductivity}\label{NCSHalleff}
In this section we will calculate both the charged and spin currents by using the extended Drude model~\cite{PhysRevLett.99.206601}. In this method the spin degree of freedom comes into play of the dynamics of charge carriers through the spin-orbital interaction, and hence  both the charged and spin currents are derived in an universal way. For the Hamiltonian \eqref{eq:nc:NonRel}, Heisenberg algebra for the canonically conjugated variables $\vec{p}$ and $\vec{x}$ gives
\begin{eqnarray}\label{eq:motionX}
    \dot{ \vec x }&=&
    \frac{\vec{p}}{m} -\frac{e \hbar}{4m^2c^2}\vec{\sigma}\times \vec{E}
    - \frac{e\sqrt{\theta} }{2}  \, \vec{E} \times \vec{x}\,,
    \\[3mm]
    \label{eq:motionP}
    \dot{ \vec p }&=&
    e\vec{E} + \frac{e\hbar}{4m^2c^2}\vec{\nabla}\big[\big(\vec{\sigma}\times\vec{E} \big)\cdot\vec{p}\big]
    - \frac{ 1 }{2} e\sqrt{\theta} \, \vec{E} \times \vec{p}\,.
\end{eqnarray}
From \eqref{eq:motionX} we have,
\begin{equation}\label{eq:ncMom}
    \vec{p} =
    m \dot{ \vec x } + \frac{ 1 }{2} e m \sqrt{\theta} \, \vec{E} \times \vec{x}
    +\frac{e \hbar}{4mc^2}  \vec{\sigma}\times \vec{E} \,,
\end{equation}
and hence
\begin{equation}\label{eq:ncAcc}
    \dot{\vec{p}} =
    m \ddot{ \vec x } + \frac{ 1 }{2} e m \sqrt{\theta} \,  \vec{E} \times \dot{\vec{x}}
    +\frac{e \hbar}{4mc^2} \big[ \dot{ \vec x } \cdot \vec{\nabla} \big] \big[ \vec{\sigma}\times \vec{E} \big]\,.
\end{equation}
Substituting \eqref{eq:ncMom} and \eqref{eq:ncAcc} into \eqref{eq:motionP}, one can get the
 dynamical equation of the canonical variable $\vec{x}$, which has the form of Newton's second law for charge carriers,
\begin{equation}\label{eq:ncNewton}
m \ddot{ \vec x } 
= \vec{F} ( e, \vec{\sigma} ) 
= \vec{F}_{e} + \vec{F}_{\theta} + \vec{F}_{\vec{\sigma}}\,.
\end{equation}
Here the ordinary Lorentz force $\vec{F}_{e} = e\vec{E}$ receives a contribution from noncommutative correction
\begin{equation}
\vec{F}_{\theta} = - e \sqrt{\theta} \, \vec{E} \times \vec{p}\,,
\end{equation}
and a spin-dependent force 
\begin{equation}
 \vec{F}_{\vec{\sigma}} =  \frac{e \hbar}{4mc^2}  \dot{ \vec x } \times \big[ \vec{\nabla} \times \big(  \vec{\sigma}\times \vec{E} \big) \big]\,,
\end{equation}
which is generated by the ordinary spin-orbital interaction,

To solve the equation \eqref{eq:ncNewton}, the velocity relaxation time
$\tau$ must be given experimentally. We assume that to the first
approximation the velocity relaxation time $\tau$ of charge carriers
is independent of $\vec{\sigma}$. Because of relative smallness of
the spin-dependent force, we can treat $\vec{F}_{\vec{\sigma}}$
 as a perturbation. The solution of  \eqref{eq:ncNewton} can be
written in the form
\begin{equation}
\dot{\vec{x}}=\dot{\vec{x}}_{e} +\dot{\vec{x}}_{\theta} +  \dot{\vec{x}}_{\vec{\sigma}}\,.
\end{equation}
Here $\dot{\vec{x}}_{e}$ is the leading order solution that is given by the ordinary Drude model~\cite{PhysRevLett.99.206601} as, 
\begin{equation}\label{eq:vLeading}
    \langle\dot{\vec{x} }_{e}\rangle=
    \frac{e\tau}{m}\vec{E}\,.
\end{equation}
For the second term $\dot{\vec{x}}_{\theta}$, which stands for the noncommutative correction, is proportional to the normal vector of the plane expanded by the electric field and the momenta along the constrained direction. Denote the unite vector in the constrained direction as $\vec{n}$, and assume that the charged particle is constrained in this direction with a length scale $\ell_{z}$, then one can easily find
\begin{equation}\label{eq:vNCC}
    \langle\dot{\vec{x} }_{\theta}\rangle=
    \frac{e\tau}{m} \frac{ \ell_{G} }{ \ell_{z} } \big( \vec{n} \times \vec{E} \big) \,,
\end{equation}
where we have used the relation $\ell_{G}=\sqrt{\theta}$. We can see that the noncommutative correction is direction-dependent, we will discuss this property latter. Finally, the spin-dependent contribution is given as 
\begin{equation}\label{FA}
    \langle\dot{\vec{x}}_{\vec{\sigma}}\rangle=
    \frac{\hbar e^2\tau^2}{4m^3c^2}
    \vec{E}\times\langle\vec{\nabla}
    \times[\vec{\sigma}\times
    \vec{E}]\rangle\,.
\end{equation}
The right-hand side of (\ref{FA}) contains the volume average of electrostatic crystal potential
$\partial_i\partial_jV_1(\vec{x})$. Furthermore, for a macroscopic system the total electric potential $V(\vec{x})$ is the sum of external electric potential $V_0(\vec{x})$ and the lattice electric potential $V_1(\vec{x})$. 
For a cubic lattice, the only invariant permitted by symmetry is~\cite{PhysRevLett.99.206601},
\begin{equation}\label{CP}
    \bigg\langle\frac{\partial^2V_1(\vec{x})}
    {\partial x_{i}\partial x_{j}}\bigg\rangle
    =\chi\delta_{ij},
\end{equation}
where $\chi$ is a constant which has been determined in Ref.~\cite{PhysRevLett.99.206601}. Then we have 
\begin{equation}\label{FA2}
    \langle\dot{\vec{x}}_{\vec{\sigma}}\rangle=
    \frac{\hbar e^2\tau^2 \chi }{2m^3c^2}
    \vec{\sigma}\times\vec{E}\,.
\end{equation}

The density matrix of the charge carriers in the spin space can be
written as,
\begin{equation}\label{SDM}
    \rho^{s}=\frac{1}{2}\rho(1+\vec{\lambda}
    \cdot\vec{\sigma}),
\end{equation}
where $\rho$ is the total concentration of charges carrying the
electric current; $\vec{\lambda}$ is  the vector of spin
polarization of the electron fluid. The total current is given by expectation value of the product between the density matrix and the velocity,
\begin{eqnarray}\label{EJ}
    \vec{j} 
    &=&
    e\langle\rho^{s} \dot{\vec{x}} \rangle
    \equiv \vec{j}^{c}+\vec{j}^{s}{\vec{\sigma}}
    (\vec{\sigma})\,,
    \\[3mm]
    \vec{j}^{c}
    &=&
    \sigma_{H}^{c}\vec{E}\,,
    \\[3mm]
    \vec{j}^{s}
    &=&
    =\sigma_{H}^{s}(\vec{\lambda}
    \times\vec{E})\,.
\end{eqnarray}
Here the corresponding Hall conductivity is given by
\begin{equation}\label{HC}
    \sigma_{H}^{c}=\frac{e^2\tau\rho}
    {m} 
    \left( \begin{array}{cc}
  1   & - \kappa_{G}   \\
\kappa_{G}  &   1   
\end{array}  \right) \,,
\end{equation}
where $\kappa_{G}=\ell_{G} / \ell_{z} $. The corresponding spin Hall conductivity does not receive correction and give as
\begin{equation}\label{SHConduct}
    \sigma_{H}^{s}=\frac{\hbar e^3\tau^2\rho
    \chi}{2m^3c^2}\,.
\end{equation}

\section{Conclusion}\label{concl}

In summary, the influences of a coordinate dependent noncommutative space on the charged and spin currents have been studied. Our results are obtained in non-relativistic limit by using the extended Drude model in Ref.~\cite{PhysRevLett.99.206601}. The non-relativistic approximation are obtained by using the
Foldy-Wouthuysen transformation, which gives general information on the nonrelativistic dynamics of the spin-1/2 particle. In case of that the noncommutativity of space is coordinate dependent, the leading correction in the Hamiltonian is a term proportional to the angular momentum of the charged particle, and independent of its spin. Therefore, only the charged current receive nontrivial contribution. The Hall conductivity receives corrections in its off-diagonal elements, while its diagonal elements remain invariant.
Therefore, the coordinate dependent noncommutativity of the space-time can be experimentally investigated by measuring the off-diagonal elements of the Hall conductivity.

\noindent\textbf{Acknowledgments}: 
K.M. is supported by the National Natural Science Foundation of China under Grant No. 11705113 and 11647018, and partially by the Shaanxi Natural Science Foundation Project under Grant No. 2017JM1032.

\bibliography{aString}

\end{document}